\documentclass[english]{article}
\usepackage[T1]{fontenc}
\usepackage[latin9]{inputenc}
\usepackage{esint}
\usepackage{babel}
\begin{document}

\title{Nonlinearity and linearity, friends or enemies? Algebraic Analyzation
of Science:)}

\author{Jerzy Hanckowiak, e-mail: hanckowiak@wp.pl , Poland, EU}

\author{(former lecturer and research worker of Wroclaw and Zielona Gora
Universities)}

\date{April, 2013}
\maketitle
\begin{abstract}
Certain operator-valued functions and new generating structures (instead
of generating functionals) are proposed for the analysis of equations
for n-point information (n-pi). Some remarks are made concerning the
intertwining of linearity and nonlinearity, and functions defined
on non-numerical objects.
\end{abstract}

\section{Introduction}

Let us start from two equations:

\begin{equation}
y=ax+b\label{eq:1}
\end{equation}
 and

\begin{equation}
y^{2}+x^{2}=r^{2}\label{eq:2}
\end{equation}
 To describe them, six different symbols were used. In the first case:
$y,=,a,x,+,b$ and in the second: $y,,2,+,x,=,r$. In the first case
we have a linear object - a straight line, in the second, we have
a nonlinear object - a circle. On these grounds, it is difficult to
decide which object is easier or more complicated to handle. In fact,
we have here used certain convention which allows to us simplify equations,
which should be written as follows:

\begin{equation}
y=a\cdot x+b\label{eq:3}
\end{equation}

\begin{equation}
\left(y\cdot y\right)^{2}+\left(x\cdot x\right)^{2}=\left(r\cdot r\right)^{2}\label{eq:4}
\end{equation}
 In this case first equation needs 7 different symbols and second
equation needs 9! In the most simple case

\begin{equation}
y=x\label{eq:5}
\end{equation}
 and

\begin{equation}
\left(y\cdot y\right)^{2}+\left(x\cdot x\right)^{2}=1\label{eq:6}
\end{equation}
 we have 3 and 8 different symbols to describe the particular straight
line and circle. If we wanted to have the same generality in a circle
as in the case of a straight line, then we would have to enter additional
2 parameters (a total of 10). These simple examples show that the
description of nonlinearity includes some additional complexity which
are not in the linear models. 

Other, well known examples, are homogeneous and non homogeneous linear
differential equations encountered in physics or engineering, to which
one can find relatively easy specific, and in many cases general solutions
or to use effective methods of approximations. These may be arguments
for trying to look for linear models, even if their original versions
are nonlinear. It turns out that this can be done at the expense of
\textbf{introducing additional, infinite number of variables} (e.g.
correlation functions), having hope that in this way the properties
of linear systems will be effectively used. Unfortunately, an infinite
number of variables needed to linearize the original nonlinear problem
leads at least to two issues: We have too many solutions which can
not be related to reasonable physical conditions, see \cite{Han },
and we have difficulties in a precise definition of a number of terms
appearing in equations for \textbf{n-point informations (n-pi)}.

These two issues were addressed in greater or lesser degree in the
previous author's papers. In the present work we will focus on a definition
of certain operator-valued functions and we introduce, for the n-pi,
instead of generating fuctionals - the new generating structures leading
to an algebraization of physics. 

In this paper we will address these two problems taking into account
linear properties of appropriate entities. At this point we would
like to note in passing that talking about linearity of considered
formulas, or equations, we usually mean that they depend on the first
power of certain \textbf{set of dependent variables}. In this way
the nonlinearity of the original theory appears into linearized theory
in different ways and in fact we are speaking about the \textbf{\textit{relative
linearity}}. 

In the case of formula (\ref{eq:3}), the \textit{absolute linearity}
would mean that

\begin{equation}
y=a+x+b\label{eq:7}
\end{equation}
which goes on to describe a stright line but inclined at an angle
of $45^{0}$ and otherwise translated with respect to the coordinate
system. This example shows that when building models of various phenomena
a linearity request should be used with a sense and we are rather
using the relative linearity, which de facto means coexistence of
non-linearity with linearity.

\section{Vector-valued functions (v-vf)}

In mathematics we are talking about linear spaces and linear mappings
(v-vf) as indeed one of the latter is closely related. Given a vector
space its elements are present as linear combinations of basis vectors
B. The numbers used in these combinations are components of vectors.
We can view this by means of (relative) linear mapping $f$:

\begin{equation}
\left\{ f(\varrho;B)\right\} _{\varrho\in R^{n}}\Longleftrightarrow V^{n}\label{eq:8}
\end{equation}
where $V^{n}$is n-D linear vector space. In fact, the linear mapping
(\ref{eq:8})

\begin{equation}
f(\alpha'\varrho'+\alpha''\varrho'';B)=\alpha'f(\varrho';B)+\alpha''f(\rho'';B)\label{eq:9}
\end{equation}
represents isomorphically n-D linear space created with vectors $\left(\varrho_{1},...,\varrho_{n}\right)$:

\begin{equation}
f(\varrho;B)\equiv\sum_{i=1}^{n}\varrho_{i}\bar{e}^{i}\Longleftrightarrow\left(\varrho_{1},...,\varrho_{n}\right)\label{eq:10}
\end{equation}
 If a new base, $B'$, is chosen, then we should have:

\begin{equation}
f(\varrho;B)=f(\varrho';B')\label{eq:11}
\end{equation}
From what we have previously said it results that for the description
of the linear vector spaces one can use v-vf depending on variables
$\varrho$ and 'parameters' $B$. 

In the case of 

\begin{equation}
f=f(\varrho;B(P))\label{eq:12-1}
\end{equation}
 we have base $B$ depending on the point $P$ and relation (\ref{eq:10})
has a local character; to every point $P$ a linear space of vectors
is related. In fact, we are dealing here with hidden nonlinearity
and, like in the linearized theories, the non-linearity does not permit
to forget about yourself. Similar intertwining of linearity and non-linearity
exists in the case of non-linear manifolds to which, at each point,
the tangent space is introduced, see also \cite{Han 2012'}.

\section*{Symmetry}

If a certain symmetry takes place, like the permutation symmetry:

\begin{equation}
\varrho=S\varrho;\; S=S^{\star}\label{eq:12}
\end{equation}
 then we have:

\begin{equation}
f(\varrho;B)=f(S\varrho;B)=f(\varrho;SB)=f(S\varrho;SB)\label{eq:13}
\end{equation}
This equality indicates that one can use a base richer than in the
absence of symmetry which may lead to left or right invertibility
of useful set of operators which make possible to introduce appropriate
projectors. 

For a linear transformation $A$

\begin{equation}
f(\varrho';B)=f(A\varrho;B)=f(\varrho;B')\Longleftrightarrow f'=\mathcal{A}f\label{eq:14}
\end{equation}
where as an excercise see: $B'=?$. 

In all these formulas the base $B$ can be finite or even uncountably
dimensional as in the case of Fock space used for description of linearized
equations, see \cite{Han 2012}. In the latter case, a linear function
$f$ depends on the uncounable number of parameters $B$.

\section{Operator-valued functions }

In many areas of science, however, non-linear functions are used which,
although they depend on one or more parameters are really difficult
to define a meaningful, because of the fact that their arguments (variables)
are operators. The \textit{functional calculus }is defined sometimes
as a branch of mathematics about inserting operators into functions
to get in result meaningful, or, at least formally correct, new operators,
see, e.g.,\cite{wiki 2012}, \cite{Hass 2007,Wiki 2010}, \cite{Han 2010,Han 2011}.
In this section we try to identify the operator associated with the
function

\begin{equation}
f(x)=a\frac{x}{1-x}\label{eq:15}
\end{equation}
using a slightly generalized functional calculus. First, we will try
to determine the operator 

\begin{equation}
f(\hat{M})=\frac{\hat{M}}{\hat{I}-\lambda_{2}\hat{M}}=?\label{eq:16}
\end{equation}
 where $\hat{M}$ is a right invertible operator. In other words,
there is an operator $\hat{M}_{R}^{-1}$ such that

\begin{equation}
\hat{M}\hat{M}_{R}^{-1}=\hat{I}\label{eq:17}
\end{equation}
 where $\hat{I}$ is the unit operator in a considered linear space
$F$, see \cite{Przew 1988} and App.3. Then in $F$ there is projector 

\begin{equation}
\hat{P}=\hat{I}-\hat{M}_{R}^{-1}\hat{M}\equiv\hat{I}-\hat{Q}\label{eq:18}
\end{equation}
projecting on the null space of the operator $\hat{M}$:

\begin{equation}
\hat{M}\hat{P}=0,\quad\hat{M}\hat{Q}=\hat{M}\label{eq:19}
\end{equation}
We would like to specify the operator (\ref{eq:16}) in such way that
the following equality would take place:

\begin{equation}
\lambda_{1}'\hat{M}\frac{1}{\hat{I}-\lambda_{2}\hat{M}}+\lambda_{1}''\frac{1}{\hat{I}-\lambda_{2}\hat{M}}\hat{M}=(\lambda_{1}'+\lambda_{1}'')\hat{B}\label{eq:20}
\end{equation}
where $\hat{B}$ is an operator given in a moment. The property (\ref{eq:20})
is weaker than the assumption that the two operators standing on the
l.h.s. of Eq.\ref{eq:20} are identical. First, we have to spacify
the formal operator 

\begin{equation}
\frac{1}{\hat{I}-\lambda_{2}\hat{M}}\equiv\hat{Y}=\left(\hat{I}-\lambda_{2}\hat{M}\right)_{R}^{-1}
\end{equation}
which we will treat as a right inverse operator to the operator $\hat{I}-\lambda_{2}\hat{M}$:

\begin{equation}
\left(\hat{I}-\lambda_{2}\hat{M}\right)\hat{Y}=\hat{I}\label{eq:22}
\end{equation}
 Multiplying this equation by the right inverse $\lambda_{2}^{-1}\hat{M}_{R}^{-1}$we
get equivalent equation:

\[
\left(\lambda_{2}^{-1}\hat{M}_{R}^{-1}-\hat{Q}\right)\hat{Y}=\lambda_{2}^{-1}\hat{M}_{R}^{-1}
\]
 hence we get the following equation for $\hat{Y}$

\begin{equation}
\left(\hat{I}-\lambda_{2}^{-1}\hat{M}_{R}^{-1}\right)\hat{Y}=\hat{P}\hat{Y}-\lambda_{2}^{-1}\hat{M}_{R}^{-1}\label{eq:23}
\end{equation}
 in which the projection $\hat{P}\hat{Y}$ of the right inverse operator
$\hat{Y}$ is an arbitrary element. Assuming, for a sake of simplicity
that $\hat{I}-\lambda_{2}^{-1}\hat{M}_{R}^{-1}$is both side invertible
operator, we get:

\begin{equation}
\hat{Y}=\left(\hat{I}-\lambda_{2}^{-1}\hat{M}_{R}^{-1}\right)^{-1}\left(\hat{P}\hat{Y}-\lambda_{2}^{-1}\hat{M}_{R}^{-1}\right)\equiv\frac{1}{\hat{I}-\lambda_{2}\hat{M}}\label{eq:24}
\end{equation}
This formula shows all the uncertainty of the expression $\frac{1}{\hat{I}-\lambda_{2}\hat{M}}$.
Now it is easy to show that if 

\begin{equation}
\hat{B}=\hat{B}\hat{Q}\label{eq:25}
\end{equation}
 and if the arbitrary term 

\begin{equation}
\hat{P}\hat{Y}=0\label{eq:26}
\end{equation}
 then Eq.\ref{eq:20} is satisfied. In this case the operator (\ref{eq:16})

\begin{eqnarray}
 & f(\hat{M})=\frac{\hat{M}}{\hat{I}-\lambda_{2}\hat{M}}\equiv\hat{B}=\hat{B}\hat{Q}\nonumber \\
 & \left\{ \hat{M}\left(\hat{I}-\lambda_{2}^{-1}\hat{M}_{R}^{-1}\right)^{-1}\left(-\lambda_{2}^{-1}\hat{M}_{R}^{-1}\right)+\left(\hat{I}-\lambda_{2}^{-1}\hat{M}_{R}^{-1}\right)^{-1}\left(-\lambda_{2}^{-1}\hat{M}_{R}^{-1}\right)\hat{M}\right\} \hat{Q}\nonumber \\
 & =\lambda_{2}^{-1}\left\{ \left(\lambda_{2}^{-1}\hat{M}_{R}^{-1}-\hat{I}\right)^{-1}+\left(\lambda_{2}^{-1}\hat{M}_{R}^{-1}-\hat{I}\right)^{-1}\hat{Q}\right\} \hat{Q}\nonumber \\
 & =2\lambda_{2}^{-1}\left(\lambda_{2}^{-1}\hat{M}_{R}^{-1}-\hat{I}\right)^{-1}\hat{Q}\label{eq:27}
\end{eqnarray}
where the projector $\hat{Q}=\hat{M}_{R}^{-1}\hat{M}$. Here it is
worth noting that the property $\hat{B}=\hat{B}\hat{Q}$, which underlies
formu\l{}u (\ref{eq:27}), is consistent with the formal expression
for the function $f$, for which $f\simeq\hat{M},$ for $\lambda_{2}\simeq0$.
Do not we will get it, if not to force a linear relationship, for
(\ref{eq:20}), with the parameters $\lambda_{1}'$ and $\lambda_{1}''$.
It is interesting, however, that B has one additional, symmetry property,
which does not have a formal prototype, namely that $\hat{B}=\hat{Q}\hat{B}$.
Could it be a clue in defining the operator valued functions?

Derived formula depends however on the choice of operator $\hat{M}_{R}^{-1}$.
Thus, additional conditions are required in order to reduce its ambiguity,
see \cite{wiki 2012}, for example, we can demand that $\hat{M}_{R}^{-1}$
is the same type as $\hat{M}$, (e.g. local), see \cite{Han 2012},
Sec.4. 

In the derivation of the above formula we was influenced mainly by
features (\ref{eq:20}) and (\ref{eq:25}) which can be only formally
justified. What does it really mean? It means so much that if the
formula (\ref{eq:16}) made sense, it would be those properties that
we want to take over the (inherited) already correctly defined formula
(\ref{eq:27}). There is another aspect here which is not insignificant
when considering the equations for n-point correlation functions,
or more general, for the n-point information (\textbf{n-pi}), namely
that the formula which is correctly specified, in many cases is the
sum of the diagonal and lower triangular operators, see previous author
papers. This means that it does not lead to additional links (branches)
with higher n-pi. Moreover, since the formal (ill defined) formula
(\ref{eq:16}) , for small $\lambda_{2}$, describes formally in many
cases polynomial interaction, we can consider the correctly defined
non-polynomial formula (\ref{eq:27}) as a candidate for description
of such polynomial interaction. Taking into account that closed equations
for n-pi can be obtained by means of highly complicated nonlinear
interactions, which approximate much simpler, polynomial interactions,
see \cite{Han' 2010 } we are inclined to say that the non-linearity
and linearity are more friends than enemies. But that last sentence
would suggest(?) that \textbf{perhaps more effective is the search
for simple equations than seeking for simple interactions!}

\section{Linearity fetish}

The popular belief is that the linearity means more simplicity and
effectiveness of the systems and phenomena description. The basic
concepts of mechanics as radius vector, force, momentum, angular momentum
- linearly depend on the variables defining them. Cartesian reference
systems include the concept of linearity for both themselves and the
relations binding them (Galilean transformation). It is widely believed
that it is easier to solve linear systems of equations than nonlinear
systems. It is surprising that it is not always the case, even when
a solution is presented in the form of formal, functional integrals
(see, e.g., quantum or stochastical field theories) because in that
case the functional integral prevents to obtain the effective final
result. Moreover, some people including me believe that the functional
integrals encountered in the field theory are generally not computable,
see \cite{Pen rose 2004}; remarks about computability. 

It turns out that the linear equations satisfied by the n-pi generated
by these functional integrals, branch out to infinity. This means
that it can not be possible to write out a reasonable, closed set
of equations, for a defined, finite set of n-pi. It is called the
\textit{closure problem }which is usually related to a nonlinearity
of the original (before linearization) equations.

\section{A new paradigm?}

These observations seem to indicate that in the existing approaches
leading to an impasse - titled the closure problem - linearity and
non-linearity are closely related. We propose to move away from the
primacy of the detailed description of the dynamic components of the
system and replacing it with the primacy of the less detailed descriptions.
In this description, in the first place, will be placed on n-pi 

\begin{equation}
<\varphi(\tilde{x}_{1})\cdots\varphi(\tilde{x}_{n})>\label{eq:28}
\end{equation}
with n=1,2,..., taking a less detailed description of the system than
description supplied by the 'field' $\varphi$, see \cite{Han 2008}.
A detailed description will be something secondary and should result
from the first in which rather global properties of the system are
taken into account. Thinking of this kind provides a not so old discoveries
in astronomy and still threatening economic crises. Climate change
also seem to suggest a different paradigm. Moving away from the detailed
description is the basis of abstraction and it allows to cope with
of extremely high complexity of the considered system. But the problem
is to use it in a more fundamental way. In this approach, concepts
such as - local, global - will play at least equivalent role. But
how to accomplish this?

\section{A 'new' approach. Free Fock space?}
\begin{description}
\item [{Motto:}]~
\end{description}
\textit{'Our experience hitherto justifies us in trusting that nature
is the realization of the simplest that is mathematically conceivable'
}Albert Einstain 

In the proposed new approach, we start with the equation on the \textit{generating
vector $|V>$ }for the functions $V(\tilde{x}_{(n)});\; n=1,2,...,\infty$:

\begin{equation}
|V>=\sum_{n=1}\int d\tilde{x}_{(n)}V(\tilde{x}_{(n)})|\tilde{x}_{(n)}>+|0>_{info}\label{eq:29}
\end{equation}
 The n-point functions $V(\tilde{x}_{(n)})$, which we call the \textit{n-point
information} (n-pi) about the system, will have different interpretation
in classical and quantum physics, see, e.g., \cite{Han' 2011}.$|\tilde{x}_{(n)}>$,
for n=1,2..., are linearly independent orthonormal vectors, the vectot
$|0>_{info}$ describes so called the \textit{local information vacuum,
}see \cite{Han 2012}. 

The generating vector $|V>$satisfies the following \textbf{linear
equation}:

\begin{equation}
\left(\hat{L}+\lambda\hat{N}+\hat{G}\right)|V>=|0>_{info}=\hat{P}_{0}|0>_{info}\label{eq:30}
\end{equation}
The operator $\hat{L}$ is a \textbf{right invertible operator}, which
in the case of \textit{classical (e.g. statistical) field theory}
is a diagonal operator:

\begin{equation}
\hat{P}_{n}\hat{L}=\hat{L}\hat{P}_{n}\label{eq:31}
\end{equation}
with respect to the projectors $\hat{P}_{n};n=1,2,...,\infty$, where
the project $\hat{P}_{n}$projects on the n-th term in the expansion
(\ref{eq:29}), see App1. The projector $\hat{P}_{0}$projects on
the subspace of the linear space F constituated by means of vectors
(\ref{eq:29}). The subspace $\hat{P}_{0}F$ does not contain any
local information about the system. In the papers \cite{Han 2012}
and \cite{Han' 2011} we call a vector belonging to subspace $\hat{P}_{0}F$
- the local information vacuum. It is surprising that both in the
classical and quantum description of systems they lead to additional
nonperturbative corrections. 

In the case of \textit{quantum field theory} the operator $\hat{L}$
is an invertible or right invertible diagonal, plus - a lower triangular
operator - related to the commutation relations of the canonical conjugate
operator variables, with respect to the same set of projectors $\hat{P}_{n}$. 

In the case of polynomial nonlinearity, the operator $\hat{N}$ is
an upper triangular operator in a classical as well as in quantum
field theory:

\begin{equation}
\hat{P}_{n}\hat{N}=\sum_{n<m}\hat{P}_{n}\hat{N}\hat{P}_{m}\label{eq:32}
\end{equation}
 see \cite{Han 2012}. 

The operator $\hat{G}$ , in the both cases, is a left invertible
operator, which is lower triangular operator:

\begin{equation}
\hat{P}_{n}\hat{G}=\sum_{m<n}\hat{P}_{n}\hat{G}\hat{P}_{m}\label{eq:33}
\end{equation}
 All these operators are linear operators acting in the \textbf{\textit{linear
space }}$F$ constituated from the vectors (\ref{eq:29}). If linearly
independent orthonormal vectors

\begin{equation}
|\tilde{x}_{(n)}>=\hat{\eta}^{\star}(\tilde{x}_{1})\cdots\hat{\eta}^{\star}(\tilde{x}_{n})|0>\label{eq:33-1}
\end{equation}
see Sec.7, then we call space $F$ the \textit{free Fock space} (FFS). 

\textbf{Why the free Fock space} (FFS) F? 

Because our experience hitherto justifies us in trusting that in such
a space constructed by means of operators satisfying Cuntz relations,
see, e.g., \cite{Han 2012}, it is easier to find the inverse operations
to multiple operators which occur in the equations for the generating
vectors as Eq.\ref{eq:30}, see previous author papers. In some sense,
we have similarity to the difference which exists in the construction
of the inverse matrices by Euler's and Gauss methods: The effectiveness
of Gauss type methods, in our opinion, is due to the fact that they
use effectively linearity of matrices themselves. In fact, this is
a FFS task. 

But that's not all. It turns out that in this space there are operators
which leads to a closed equation for n-pi, see \cite{Han 2012,Han' 2010 ,Han 2010},
and so on, that for small values of the so called minor coupling constant
at least formally approximate operators used in the \textbf{usual
(not free) Fock space}.  

It is not excluded that in this way the entangled together problems
of non-linearity and closuring are significantly overcome. 

In FFS it is also possible to introduce vectors describing local and
global information which allows the use of a unique language to describe
phenomena belonging to different fields of human activity such as
physics or economics, complex systems, etc.

\section{New generating structures describing classical and quantum physics;
noncommutative rings and algebraization of physics}

In fact, we can avoid the introduction of the vector space of generating
vectors (\ref{eq:29}) by considering only (generating) operators
acting in FFS. For this purpose, instead of vectors (\ref{eq:29})
with the base vectors $|\tilde{x}_{(n)}>=\hat{\eta}^{\star}(\tilde{x}_{1})\cdots\hat{\eta}^{\star}(\tilde{x}_{n})|0>$,
we introduce the lower triangular operator 

\begin{equation}
\hat{V}_{0}\equiv|V><0|=\sum_{n=1}\int d\tilde{x}_{(n)}V(\tilde{x}_{(n)})\hat{\eta}^{\star}(\tilde{x}_{1})\cdots\hat{\eta}^{\star}(\tilde{x}_{n})\hat{P}_{0}+\hat{P}_{0}\label{eq:34}
\end{equation}
 with $\tilde{x}_{(n)}\equiv\tilde{x}_{1},...,\tilde{x}_{n}$ and
$\hat{P}_{0}=|0><0|$, where operators $\hat{\eta}$satify the Cuntz
relations

\begin{equation}
\hat{\eta}(\tilde{y})\hat{\eta}^{\star}(\tilde{x})=\delta(\tilde{y}-\tilde{x})\cdot\hat{I}\label{eq:35}
\end{equation}
and vectors $|0>,<0|$ describe local information vacuum, see \cite{Han 2012}.
The star means an involution of the operator $\hat{\eta}$, $\left(\hat{\eta}^{\star}\right)^{\star}=\hat{\eta}$,
and the projector 

\begin{equation}
\hat{P}_{0}\sim\hat{P}_{info}\label{eq:36}
\end{equation}
 We also assume that, for arbitrary 'vectors' $\tilde{x}$, the projector
$\hat{P}_{0}=\hat{P}_{0}^{\star}$ and operators $\hat{\eta}$ have
the following properties: 

\begin{equation}
\hat{\eta}(\tilde{x})\hat{P}_{0}=\hat{P}_{0}\hat{\eta}^{\star}(\tilde{x})=0\label{eq:37}
\end{equation}
 From the above, we have

\begin{equation}
\hat{P}_{0}\hat{\eta}(\tilde{y}_{1})\cdots\hat{\eta}(\tilde{y}_{n})\hat{V}_{0}=V(\tilde{y}_{(n)})\hat{P}_{0}\label{eq:37-1}
\end{equation}
 We also have

\begin{equation}
\hat{V}_{0}=\hat{V}_{0}\hat{P}_{0},\;\hat{P}_{0}\hat{V}_{0}=\hat{P}_{0}\label{eq:37-2}
\end{equation}

Operators $\hat{V}_{0}$ satisfy very similar equation as Eq.\ref{eq:30}:

\begin{equation}
\left(\hat{L}+\lambda\hat{N}+\hat{G}\right)\hat{V}_{0}=\hat{P}_{info}\sim\hat{P}_{0}\label{eq:38}
\end{equation}

One can introduce a more general generating operators than the operators
(\ref{eq:34}), with diagonal, lower and upper triangular elements,: 

\begin{equation}
\hat{V}=\sum_{n=0}^{\infty}\sum_{m=0}^{\infty}\int d\tilde{x}_{(m)}d\tilde{y}_{(n)}V_{m,n}(\tilde{x}_{(m)},\tilde{y}_{(n)})\hat{\eta}^{\star}(\tilde{x}_{1})\cdots\hat{\eta}^{\star}(\tilde{x}_{m})\hat{P}_{0}\hat{\eta}(\tilde{y}_{1})\cdots\hat{\eta}(\tilde{y}_{n})+\hat{P}_{info}\label{eq:39}
\end{equation}
while we agree that the subscript zero means that the variable does
not exist in the given expression. 

We have:

\begin{equation}
\hat{P}_{0}\hat{\eta}(\tilde{x}_{1})\cdots\hat{\eta}(\tilde{x}_{m})\hat{V}\hat{\eta}^{\star}(\tilde{y}_{1})\cdots\hat{\eta}^{\star}(\tilde{y}_{n})\hat{P}_{0}=V_{m,n}(\tilde{x}_{(m)},\tilde{y}_{(n)})\hat{P}_{0}\label{eq:40}
\end{equation}
 for $k=0,...,n$ and $n=0,1,...,\infty$. 

We postulate, for the operators $\hat{V}$, the following equation:

\begin{equation}
\left(\hat{L}+\lambda\hat{N}+\hat{G}\right)\hat{V}=\hat{\Phi}\label{eq:41}
\end{equation}
with a 'source' operator $\hat{\Phi}$ . Imposing on$\hat{V}$ the
condition:

\begin{equation}
\hat{P}_{0}\hat{V}=\hat{P}_{0}\label{eq:42}
\end{equation}
and on the source term $\hat{\Phi}$ the condition:

\begin{equation}
\hat{\Phi}\hat{P}_{0}=\hat{P}_{info}\label{eq:43}
\end{equation}
we see that the component $\hat{V}\hat{P}_{0}\equiv\hat{V}_{0}$of
the generating operator $\hat{V}$ satisfies exactly the same Eq.\ref{eq:38}
as the generating operator $\hat{V}_{0}$. 

An algebraization of equations introduced here leads to description
of considered equations in which the sought entities and entities
used to describe equations - belong to the same category of notions,
for example, they are operators. This allows for raising new questions,
To see this let us assume that the generating operator $\hat{V}$
satisfies a more general equation

\begin{equation}
\hat{A}\hat{V}=\hat{\Phi}\label{eq:44}
\end{equation}
with given operators $\hat{A},\hat{\Phi}$. Now, in addition to Eq.\ref{eq:42}
we postulate that

\begin{equation}
\hat{\Phi}\hat{P}_{0}=\hat{\Phi}_{0}\label{eq:44-1}
\end{equation}
 which may be different from Eq.\ref{eq:43}. 

Like in the case of Eq.\ref{eq:30}, let us assume that the operator
$\hat{A}$ is a right invertible. This means that a solution can be
expressed as

\begin{equation}
\hat{V}=\hat{A}_{R}^{-1}\hat{\Phi}+\hat{P}_{A}\hat{V}\label{eq:45}
\end{equation}
with an arbitrary projection $\hat{P}_{A}\hat{V}$, where $\hat{P}_{A}=\hat{I}-\hat{A}_{R}^{-1}\hat{A}$
is a projector on the null space of the operator $\hat{A}$ and $\hat{A}_{R}^{-1}$
is its a right inverse. Now we can see what we would get if the generating
operator $\hat{V}$ was also a right invertible with $\hat{V}_{R}^{-1}$
as its right inverse: $\hat{V}\hat{V}_{R}^{-1}=\hat{I}$? From Eq.\ref{eq:45}
, we get 

\begin{equation}
\hat{I}=\hat{A}_{R}^{-1}\hat{\Phi}\hat{V}_{R}^{-1}+\hat{P}_{A}\label{eq:46}
\end{equation}
 and this would mean that product of operators 

\begin{equation}
\hat{A}_{R}^{-1}\hat{\Phi}\hat{V}_{R}^{-1}=\hat{I}-\hat{P}_{A}\equiv\hat{Q}_{A}\label{eq:47}
\end{equation}
 which is a projector, would not depend on the arbitray source operator
$\hat{\Phi}$. But on the above limitation one can look in a more
positive way, namely, that in the case of a more fundamental theory
the 'sources' (including in this name the currents) and interactions
described by operators $\hat{\Phi}$ and $\hat{A}$ are somehow related
to each other. In fact, Eq.\ref{eq:47} like original Eq.\ref{eq:44}
relates three entities: $\hat{A}_{R}^{-1},\hat{\Phi}$and $\hat{V}_{R}^{-1}$.
But in the case of (\ref{eq:47}) this relation is a more restrictive
and indicating a certain 'entanglement' or unification of them. 

Multiplying Eq.\ref{eq:47} by $\hat{A}$ we get equation

\begin{equation}
\hat{\Phi}\hat{V}_{R}^{-1}=\hat{A}\hat{Q}_{A}=\hat{A}\label{eq:48}
\end{equation}
 which can be regarded as an equation for a right inverse $\hat{V}_{R}^{-1}$.
Having calculated $\hat{V}_{R}^{-1}$, we can calculate the operator
$\hat{V}^{\star}$by means of the equation:

\begin{equation}
\left(\hat{V}_{R}^{-1}\right)^{\star}\hat{V}^{\star}=\hat{I}\label{eq:49}
\end{equation}
 and the generating operator $\hat{V}=\left(\hat{V}^{\star}\right)^{\star}$.
Hence, finally, 

\begin{equation}
|V>=\hat{V}|0>\label{eq:50}
\end{equation}
It shows how algebraization of equations allows for a new approache
to old problems. 

In the case of transformed Eq.\ref{eq:30}, and after its symmetrization,
see, e.g.,\cite{Han 2012},

\begin{eqnarray}
 & \left(\hat{I}+\lambda\left(\hat{I}+\hat{S}\hat{L}_{R}^{-1}\hat{G}\right)^{-1}\hat{S}\hat{L}_{R}^{-1}\hat{N}\right)|V>=\nonumber \\
 & \left(\hat{I}+\hat{S}\hat{L}_{R}^{-1}\hat{G}\right)^{-1}\left(\hat{S}\hat{L}_{R}^{-1}|0>_{info}+\hat{S}\hat{P}_{L}|V>\right)\label{eq:51}
\end{eqnarray}
 we can consider the operator equation

\begin{equation}
\left(\hat{I}+\lambda\left(\hat{I}+\hat{S}\hat{L}_{R}^{-1}\hat{G}\right)^{-1}\hat{S}\hat{L}_{R}^{-1}\hat{N}\right)\hat{V}=\left(\hat{I}+\hat{S}\hat{L}_{R}^{-1}\hat{G}\right)^{-1}\left(\hat{S}\hat{L}_{R}^{-1}\hat{P}_{info}+\hat{S}\hat{P}_{L}\hat{V}\right)\label{eq:52}
\end{equation}
 This means that in Eq.\ref{eq:44}, the operator

\begin{equation}
\hat{A}=\left(\hat{I}+\lambda\left(\hat{I}+\hat{S}\hat{L}_{R}^{-1}\hat{G}\right)^{-1}\hat{S}\hat{L}_{R}^{-1}\hat{N}\right)\label{eq:53}
\end{equation}
 and the operator 

\begin{equation}
\hat{\Phi}=\left(\hat{I}+\hat{S}\hat{L}_{R}^{-1}\hat{G}\right)^{-1}\left(\hat{S}\hat{L}_{R}^{-1}\hat{P}_{info}+\hat{S}\hat{P}_{L}\hat{V}\right)\label{eq:54}
\end{equation}

\section{Final remarks and comments on symmetrization of calculations; too
much symmetry in science?}

The operator-valued functions of the right invertible operators incorporating
three properties of the formal formula as:

i. linearity, see Eq.\ref{eq:20},

ii. commutativity with respect to the $\hat{M}$ operator, see again
Eq.\ref{eq:20}, and 

iii. projecton properties, see Eq.\ref{eq:27} 

where constructed. 

To construct such operator-valued functions, we did not take into
account spectral properties of used operators like in the usual approach
of the functional calculus, but we have used the most primitive methods
of defining such functions, namely, to present them in the form of
infinite power series. The specificity of the submitted approach is
that in many cases, for interesting projections, only a finite number
of terms of such series gives contributions , see previous author
papers. In this sense a new approach, illustrated by the motto to
Sec.6, is possible:). 

An important element of this paper is also a new generating structure
(operator) for the n-pi $V(\tilde{x}_{(n)})$ that allows to describe
systems and considered equations by means of the \textbf{noncommutative
ring with the unity}, see, e.g., \cite{Lidl 1983}. In this way, the
obstacle has been removed associated with the use of the basic concepts
of physics, namely - vector spaces in which are not defined vector
products. In the proposed approach has been abolished demarcation
line between the description of equations (operators) and the description
of physical systems (vectors). For both objects, we use the elements
of noncommutative ring with unity. 

Usually, the division on the operators and vectors is justified \textbf{by
the demand} that an action of the operator on the vector should gives
the vector. Such deman is automatically realized by the ring in which
vectors are represented by the one column matrices, first one. If,
however, we resigne from that demand then the vectors can be substituted
by the operators, e.g. the diagonal matrices. In many cases such matrices
representing vectors can be inverted which is a useful property in
many solving procedures. 

Similar reasoning lies behind the idea of replacing the generating
functions or functionals by the generating vectors. In this case,
it was possible due to the fact that the generating functions or functionals
do not have to be covergent. As a result, obtained equations admit
more general representations. 

And one more thing related to the paper title: the considered generating
structures depend in the nonlinear way on the auxiliary field operators
$\hat{\eta}$ and $\hat{\eta}^{\star}$, see, for example, formula
(\ref{eq:34}). One can introduce the equivalent generating structures
which linearly depend on the infinite set of auxiliary n-point functions
$\varrho(\tilde{x}_{(n)})$, see \cite{Han 2012}. This leads, as
we think, to the more complex formulas, especially in defining the
operator-valued functions. 

As far as the algebraization idea of physics and science in general,
we would like to note that it is much less appreciated by the scientific
community than the geometrization idea. In fact, we think that \textbf{there
is too much symmetry in science} which is reflected in the assumptions
on the generating fuctions or functionals - the entities having only
auxiliary character. The mere transfer of symmetry of physical quantities
on the generating structures leads in general to the divergent power
series which we call the formal power series. Excess of symmetry is
often masked in a natural way as a result of differences in the laws
of nature (equations) from the initial or/and boundary conditions.
We speak then about \textbf{spontaneous symmetry breaking}. You must
also be aware of the fact that the very existence of the reference
frame disturbs the symmetry of the described system, for instance
the Universe. see \cite{Ros 2008}. See also \cite{Shea 2013}. 

Each symmetry is associated with some limitations. So if the auxiliary
entities will unnecessarily inherit restrictions that apply to the
physical entities then we will needlessly deprive ourselves \textbf{effective
calculations}. Since the multiscale, complex systems mostly deal with
permutation symmetry, it is worth recalling the \textbf{Cayley's theorem}
that every group is isomorphic to a group of permutations. See also
\textbf{Klein's Erlangen program }of relation of symmetry with geometry\textbf{. }

That's what we're talking about is similar to reductionism in science,
which uses a quasi-invariance (quasi-isolation) of the system under
study with respect to changes in the environment. In this analogy,
the environment would be a generating vector or operator. \textbf{Reduction
is symmetry}, see \cite{Ros 2008}. 

We believe that algebraization of description of the multiscale and
complex systems will significantly improve the process of computing.
It will also allow for a broader look at the different areas of mathematics
and physics, see \cite{Edwa 1993,Yad 2001,Han 2012'}, and especially
\cite{Hell 2012}. See, however, \cite{Ros 2008}. See also \cite{Wall 2010},
where algebraic approach to quantum field theory is criticized, but
there it was mainly concerned with the problem of renormalization. 

At the end of work I would like to draw attention to the fact that
the \textbf{description of physical systems} based on moving away
from the details and its algebraization leading to the noncomutative
rings is similar to the way that leads to free probability and noncommutative
geometry, see \cite{Hell 2012}. The difference lies in the fact that
approach proposed here is realized in a more transparent manner.

We would also like to draw the reader's attention to another aspect
of the generalization of this and previous works. In Eq.\ref{eq:30}the
term associated with the \textbf{linear nature of the phenomenon},
the so-called kinematic term, is described by the operator $\hat{L}$,
which is the right invertible operator. In the simplest case, this
would be a derivative of the first or higher orders. Maybe this is
the reason why sometimes right invertible operators are called the
\textit{derivatives}. To these operators are related the basic physical
quantities such as velocity, acceleration, the existence of free waves
and the existence of physically interesting solutions to considered
equations. Interesting is also the fact that these are mostly diagonal
plus lower triangular operators. \textbf{Nonlinear phenomena }are
described mostly by the upper trangular operators, and external fields
in which systems are submerged, are described by the lower triangular,
left invertible operators. By the lower triangular operators are also
described quantum properties of systems, see \cite{Han 2011-1}. For
this reason, it seems interesting definition of operator-valued functions
considered in Sec.3 which is not leaving the above class of operators
at least, for polynomial nonlinearity. Does this mean a greater unification
of linear and non-linear, or, classical and quantum phenomena? That
is the question.

\section*{App.1 Projectors $\hat{P}_{n}$.}

These are projectors projecting on the n-th terms of the expansion
(\ref{eq:34}). They are:

\begin{equation}
\hat{P}_{n}=\int\hat{\eta}^{\star}(\tilde{x}_{1})\cdots\hat{\eta}^{\star}(\tilde{x}_{n})\hat{P}_{0}\hat{\eta}(\tilde{x}_{n})\cdots\hat{\eta}(\tilde{x}_{1})d\tilde{x}_{(n)}\label{eq:55}
\end{equation}
 for n=1,2,.... They can be expressed in another form as:

\begin{equation}
\hat{P}_{n}=\int\hat{\eta}^{\star}(\tilde{x}_{1})\cdots\hat{\eta}^{\star}(\tilde{x}_{n})\left(\hat{I}-\int\hat{\eta}^{\star}(\tilde{x})\hat{\eta}(\tilde{x})d\tilde{x}\right)\hat{\eta}(\tilde{x}_{n})\cdots\hat{\eta}(\tilde{x}_{1})d\tilde{x}_{(n)}\label{eq:56}
\end{equation}
 in which the \textit{vacuum projector} $\hat{P}_{0}$does not appear.
The name of $\hat{P}_{0}$ comes from interpretation of functions
$V(\tilde{x}_{(n)})$ as the n-p-i about the field $\varphi$, see
(\ref{eq:28}). We have got, of course, that the unit operator $\hat{I}$,

\section*{
\begin{equation}
\hat{I}=\sum_{n=1}^{\infty}\hat{P}_{n}+\hat{P}_{0}\label{eq:57}
\end{equation}
 App.2 Other projectors}

The operator

\begin{equation}
\hat{R}=\sum_{n=1}\int d\tilde{x}_{(n)}\hat{\eta}^{\star}(\tilde{x}_{1})\cdots\hat{\eta}^{\star}(\tilde{x}_{n})\hat{P}_{0}+\hat{P}_{0}\equiv\sum_{n=1}\hat{R}_{n}+\hat{P}_{0}\label{eq:58}
\end{equation}
and operators $\hat{R}_{n}$ are very lower triangular operators with
respect to projectors $\hat{P}_{n}$. We can see that

\begin{equation}
\hat{R}_{n}^{\star}\hat{R}_{n}=\hat{P}_{0}vol^{n}\label{eq:59}
\end{equation}
 and

\begin{equation}
\hat{R}_{n}\hat{R}_{n}^{\star}\cdot\hat{R}_{n}\hat{R}_{n}^{\star}=\hat{R}_{n}\hat{R}_{n}^{\star}\cdot vol^{n}\label{eq:60}
\end{equation}
 In other words, the diagonal products $\hat{R}_{n}\hat{R}_{n}^{\star}$behave
like pseudo-projectors which for the unit volume are projectors. In
fact they are projectors after division by $vol^{n/2}$. In contrast
to the orthogonal projectors $\hat{P}_{n}$:

\begin{equation}
\hat{P}_{m}\hat{P}_{n}=\delta_{mn}\hat{P}_{n}\label{eq:61}
\end{equation}
 projectors $\hat{R}_{n}$are not orthogonal.

\section*{App.3 Algebraic analysis?}

In this as well as in the previous papers we are using certain results
of algebraic analysis, see \cite{Przew 1988}. Since the same name
stands for two different branches of mathematics, see \cite{Przew 2000}
to form an opinion on this terminological confusion.

\end{document}